# Broadband THz spectroscopy of the insulator-metal transition driven by coherent lattice deformation at the SmNiO$_3$/LaAlO$_3$ interface

W. Hu[1,*], S. Catalano[2], M. Gibert[2], J.-M. Triscone[2], and A. Cavalleri[1,3,†]

1. Max Planck Institute for the Structure and Dynamics of Matter, Center for Free Electron Laser Science, 22761 Hamburg, Germany
2. Department of Quantum Matter Physics, University of Geneva, CH-1211 Geneva, Switzerland
3. Department of Physics, Clarendon Laboratory, University of Oxford, OX1 3PU Oxford, United Kingdom

**Abstract:** We investigate the non-equilibrium insulator-metal transition driven in a SmNiO$_3$ thin film by coherent optical excitation of the LaAlO$_3$ substrate lattice. By probing the transient optical properties over a broad frequency range (100 - 800 cm$^{-1}$), we analyze both the time dependent metallic plasma and the infrared optical phonon line shapes. We show that the light-induced metallic phase in SmNiO$_3$ has the same carrier density as the equilibrium metallic phase. We also report that the LaAlO$_3$ substrate acts as a transducer only at the earlier time delays, as the vibrations are driven coherently. No long-lived structural rearrangement takes place in the substrate. Finally, we show that the transient insulator-metal transition occurs both below and above the Néel temperature. We conclude that the supersonic melting of magnetic order measured with ultrafast x-rays is not the driving force of the formation of the metallic phase. We posit that the insulator metal transition may origin from the rearrangement of ordered charges at the interface propagating into the film.

[*] wanzheng.hu@mpsd.mpg.de
[†] andrea.cavalleri@mpsd.mpg.de

The metal-insulator and magnetic transitions of nickelates [1,2] can be systematically manipulated at equilibrium with interfacial strain. By changing substrate materials and introducing either compressive or tensile strain one can for example tune the metal-insulator transition temperature over several hundred Kelvins for the nickelate films.

In the same spirit, we have shown in the past that by optically exciting the lattice of $LaAlO_3$ or $NdGaO_3$ substrates, one could induce ultrafast switching of the electrical and magnetic properties of the nickelate thin film $NdNiO_3$. A long-lived, five-order-of-magnitude increase in the low-frequency conductivity [3] was measured, accompanied by the propagation of a supersonic antiferromagnetic/paramagnetic front [4]. The relationship between substrate excitation, creation of mobile charges, melting of magnetism and a possible structural rearrangement in the nickelate film is yet to be determined.

Firstly, to date we could not determine if the transient state reaches the same metallic phase as that achieved thermally at equilibrium. Because the transient reflectivity was probed over a limited frequency range (up to 25 meV, 200 cm$^{-1}$) [3], we could not cover the whole Drude peak in $NdNiO_3$ [5] and the phonon modes in the $LaAlO_3$ substrate [6]. Hence, the total Drude spectral weight in the transient phase could not be retrieved reliably. The experiments reported here show that the nickelate film switches to a state that has the same Drude weight of the equilibrium metallic phase. This effect is however also established to be not thermal, as the substrate structural peaks reflect minimal heating.

Secondly, the role of the substrate lattice in switching the electronic phase in the thin film has been unknown. Indeed, it has not been clear if the lattice of the substrate is undergoing a long-lived structural rearrangement that sustains the metallic phase, or if the coupling is only taking place at early times. In the data reported here we show that no rearrangement takes place in the substrate and that the phonon resonances are not modified after the interaction with the pump light.

Thirdly, to date it was not clear if the melting of magnetism observed with ultrafast x-ray scattering [4] causes the light-induced insulator-metal transition. This cannot be determined easily in NdNiO$_3$ because in this material the insulator-metal transition and the antiferromagnetic ordering temperatures coincide [7]. By measuring in SmNiO$_3$/LaAlO$_3$, which exhibits well-separated metal-insulator transition and Néel temperatures [7], we show here that the transient insulator-metal transition can be induced both in the antiferromagnetic and paramagnetic insulators, and that melting of magnetism observed in NdNiO$_3$ [4] is not a necessary condition for the ultrafast insulator-metal transition.

SmNiO$_3$ epitaxial thin film (24 unit cells, 10 nm thick) were used, grown by off-axis radio frequency magnetron sputtering on a (001) LaAlO$_3$ single crystal substrate [2]. The LaAlO$_3$ substrate was a 5x5 mm$^2$ size, twinned sample with edges parallel to the pseudocubic crystallographic axes.

The equilibrium optical and transport properties of the nickelate heterostructure are shown in Fig. 1. The in-plane reflectivity of the bare substrate and the nickelate heterostructure were measured at near-normal incidence from 100 to 15000 cm$^{-1}$ using a Bruker vertex 80v spectrometer. A gold mirror was used as a reference for the reflectivity measurement at various temperatures from 10 to 295 K. As plotted in Fig. 1(a), the reflectivity of the bare substrate is dominated by three square-shaped peaks around 200, 450 and 700 cm$^{-1}$ in the far-infrared region. These are infrared-active phonons [8] with strong temperature dependence. In particular, the maximum reflectivity at the 700 cm$^{-1}$ phonon decreases from 0.75 to 0.6 when increasing the temperature from 10 to 295 K. Figure 1(b) is the far-infrared reflectivity of SmNiO$_3$/LaAlO$_3$. The reflectivity resembles that of the bare substrate, since the overall light penetration depth at far-infrared (~μm) is much deeper than the SmNiO$_3$ film thickness (10 nm). The rounding off of the flat substrate phonon bands is due to a finite conductivity in SmNiO$_3$, including the free-carrier contribution and the low-frequency tail of the insulating and charge-order gaps [9]. Our optical spectra agree very well with earlier data on SmNiO$_3$/LaAlO$_3$ [9, 10]. The insulator-metal transition temperature for our SmNiO$_3$/LaAlO$_3$ sample is 370 K from the four-probe resistivity measurement, and the antiferromagnetic ordering temperature is 200 K revealed by resonant X-ray diffraction (Fig. 1(c)) [2].

For the pump-probe experiment, we used mid-infrared pump pulses to drive the substrate phonon, and measured the transient broadband reflectivity at normal incidence with linearly polarized light along the [100] direction [11]. The broadband probe pulses were generated and probed in a laser-ionized gas plasma [12], using 800 nm pulses with 1 mJ energy and 35 fs duration from a Ti:sapphire laser. To measure the high frequency range with sufficient signal-to-noise ratio, the probe pulses were further detected in a 50-µm-thick Z-cut GaSe crystal, which was tilted ~45 degrees with respect to the incident beam [13]. In this way, we could measure the transient reflectivity from 300 to 800 cm$^{-1}$, which is crucial for detecting possible structural changes of the infrared-active phonons in LaAlO$_3$.

The mid-infrared pump pulses were generated by difference-frequency mixing in a home-built two-stage optical parametric amplifier. The center frequency was tuned to the 750-cm$^{-1}$ longitudinal plasma frequency of the oxygen stretching mode [8] in LaAlO$_3$. Because of the twinned structure of the rhombohedral LaAlO$_3$ and the strong mixing of the phonon eigenvectors in perovskites [14], our pump pulses drove the oxygen stretching mode with mixed A$_{2u}$/E$_u$ symmetries. The time duration of the pump pulses is 300 fs. The pump fluence is 2 mJ/cm$^2$.

We first studied the transient reflectivity of the bare substrate at $T$ = 10 K. Figure 2(a) plots the equilibrium and transient reflectivity at selected time delays. The transient reflectivity changes, $\Delta R = R^{\text{transient}} - R^{\text{equilibrium}}$, from 0 to 30 ps are shown as color plots in Fig. 2(b) and (c). No detectable phonon reshaping or intensity changes can be seen at all measured time delays, indicating the lattice symmetry remains the same as the equilibrium one. Excitation of the substrate lattice, which follows the high frequency mid-infrared field during the first 150 fs, does not change on average. By comparing the measured data with the strong temperature dependence of the reflectivity (Fig. 1(a)), we conclude that light-induced heating is negligible.

We measured the transient reflectivity of the SmNiO$_3$/LaAlO$_3$ heterostructure with the pump pulses polarized either perpendicular or parallel to the probe pulses. In the pump⊥probe case, a further rounding off of the substrate phonon bands with increasing time delays can be seen in Fig. 2(d) and the differential reflectivity plots (Fig. 2(e) and (f)). The transient reflectivity increases (blue)/decreases (red) below/above the transverse plasma frequencies of the substrate phonon modes at 450 and 700 cm$^{-1}$. The transient reflectivity change reaches maximum at 1 ps time delay.

Since experiments in the bare LaAlO$_3$ substrate yields no pump-probe response (Fig. 2(a)-(c)), the reflectivity changes reported here for the LaAlO$_3$/SmNiO$_3$ should come from the nickelate thin film. When the pump pulses were polarized parallel to the probe pulses, we observed a transient rounding off of the substrate phonon bands comparable to that seen in the pump⊥probe case. However, a new absorption feature emerged around 700 cm$^{-1}$ (Fig. 2 (i)), which is close to the transverse plasma frequency of the substrate oxygen stretching mode. This 700-cm$^{-1}$ feature develops after the pump pulses have left, red-shifts and then blue-shifts, and finally disappears before 1 ps, the time when the transient rounding off of the substrate phonon bands reaches maximum.

The transient reflectivity reported above was fitted with the same model used for the material at equilibrium. In both cases, we fitted the substrate and the thin film separately by Drude-Lorentz terms, and then used a two-layer model [15] to fit the reflectivity of the heterostructure,

$$\tilde{r}_{SmNiO3/LaAlO3} = \frac{\tilde{r}_{SmNiO3} + \tilde{r}_{LaAlO3}exp(2i\delta)}{1 + \tilde{r}_{SmNiO3}\tilde{r}_{LaAlO3}exp(2i\delta)}$$

where $\delta = 2\pi d(n_{SmNiO3}+ik_{SmNiO3})/\lambda$. Here $n_{SmNiO3}$ is the refractive index and $k_{SmNiO3}$ is the extinction coefficient of the SmNiO$_3$ layer, $d$ is the film thickness, and $\lambda$ is the wavelength of the probe pulse. $\tilde{r}_{SmNiO3/LaAlO3}$, $\tilde{r}_{SmNiO3}$ and $\tilde{r}_{LaAlO3}$ are the reflection coefficients of the heterostructure, the SmNiO$_3$ layer, and the LaAlO$_3$ substrate, respectively.

Because measurements in the bare substrate did not show any light-induced changes, we used the equilibrium fitting parameters for the substrate, and time-varying parameters for the nickelate layer to fit the transient reflectivity. The fitting was performed under the assumption of a homogeneous response for the thin film.

For the pump⊥probe case, the transient reflectivity at all time delays was well fitted by changing only the zero-frequency conductivity value in the nickelate layer; for the pump//probe case, extra Lorentzians are used for the nickelate layer to fit the 700-cm$^{-1}$ feature at early delays (Fig. 2(i)). In the end, we obtained similar zero-frequency conductivity values for the parallel and the perpendicular pump-probe geometries.

Figure 3(a) shows the fitting result for the broadband transient reflectivity of SmNiO$_3$/LaAlO$_3$ at 1 ps and $T$ = 10 K. The metallic state conductivity value σ$_{dc}$($T$ = 400 K) was used to fit the SmNiO$_3$ layer, while all phonon parameters were kept unchanged for the LaAlO$_3$ substrate. Similar to the equilibrium case [10], SmNiO$_3$ is a bad metal in the transient state. The real part of the optical conductivity shows a broad Drude peak with a scattering rate $1/\tau$ = 1000 cm$^{-1}$ (Fig. 3(b)). Similar fits were done for other time delays. The extracted zero-frequency conductivity of SmNiO$_3$ is shown as dark-blue squares in Fig. 3(c). The data at -1 ps is the equilibrium value, calculated from the resistivity (Fig. 1(c)). The error bars are standard deviations from various experimental runs. The mean value of the transient conductivity increases to 2190 Ω$^{-1}$cm$^{-1}$ within 1.5 ps, and then decays exponentially back to the equilibrium value. According to the Drude model, the zero-frequency conductivity is $\sigma_{dc} = ne^2\tau/m^*$, where $n$ is the carrier density, $1/\tau$ is the scattering rate and m* is the carrier effective mass. A constant scattering rate was used to fit all pump-probe time delays. Assuming that the carrier effective mass does not change in the light induced state, we relate the increase in the zero-frequency conductivity to an increase in carrier density.

Figure 3(c) displays the fitted transient conductivity for different base temperatures. The maximum transient conductivities reach the same level within the experimental error, independent of whether the starting equilibrium phase is an antiferromagnetic insulator ($T$ = 10 and 150 K) or a paramagnetic insulator ($T$ = 230 and 295 K). Consistently, the life-time of the transient state did not show any correlation with the antiferromagnetic ordering temperature (Fig. 3(d)), increasing continuously with increasing temperature and smoothly crossing the Néel temperature.

Because neither the transient conductivity or the lifetime of the transient state show anomalies at the equilibrium Néel temperature, we conclude that the light induced effect proceeds independently of pre-existing magnetic order. The physics of the interface control may then be driven either by a structural phase transition launched in the material or by a purely electronic mechanism, for example fast melting of charge order. The present data cannot resolve this problem.

Our data offer new insight into understanding the role of the driven substrate lattice in triggering the insulator-metal transition in the nickelate. No phonon splitting or frequency shifting for the substrate phonons is observed in the transient state, which rules out possible symmetry changing in the substrate. Contrary to the 20% reflectivity drop around 700 cm$^{-1}$ with increasing temperature (Fig. 1(a)), no observable laser heating in the substrate (Fig. 2(a-c)), which excluded the possibility that a thermally-induced strain modification may trigger the insulator-metal transition in the nickelate. The reflectivity reshaping at the transverse plasma frequency of the driven substrate phonon indicates the substrate as a transducer only at early time delays, driving the transient insulator-metal transition in the SmNiO$_3$ film.

In summary, we have demonstrated a transient conductivity enhancement in the insulating $SmNiO_3$ thin film when the substrate phonon is driven by mid-infrared pulses. This is reminiscent of the effect observed in $NdNiO_3$ films [3]. The transient state in $SmNiO_3$ has the same carrier density as the equilibrium high-temperature metallic phase, despite negligible heating. The maximum transient conductivity reaches the same metallic-state value, starting from the antiferromagnetic and paramagnetic insulating phases. This indicates that the light-induced insulator-metal transition is not driven by the melting of the magnetic order. More generally, our work offers new insight into understanding and further exploring the control of hetero-interfaces with light.

ACKNOWLEDGMENTS

We thank Stefan Kaiser, Roberto Merlin, Miroslav Abrashev, Stefan Zollner, Alaska Subedi, Oleg E. Peil, Roman Mankowsky and Michael Först for extensive discussions. The research has received financial support from the European Research Council under the European Union's Seventh Framework Programme (FP7/2007-2013)/ERC Grant Agreement no. 319286 (Q-MAC). Part of this work was supported by the Swiss National Science Foundation through Division II.

FIGURES

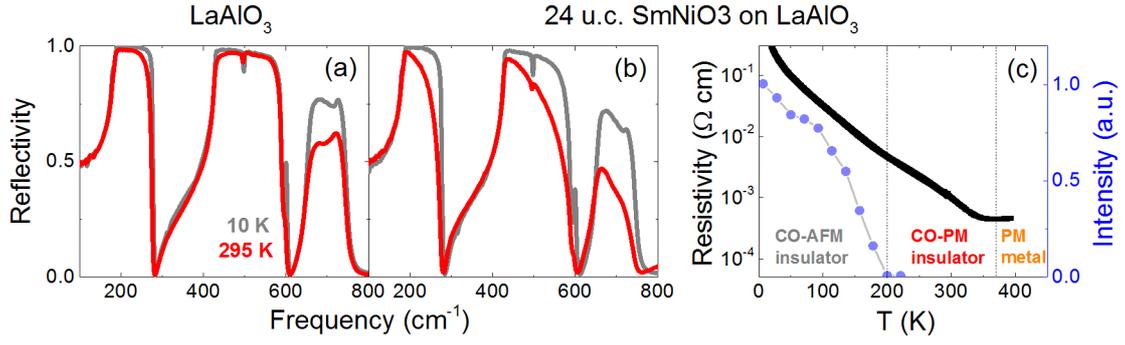

FIG. 1. (Color online) Equilibrium optical reflectivity of (a) the bare LaAlO$_3$ substrate and (b) the nickelate heterostructure. For both materials, the optical response is dominated by three infrared-active phonons of LaAlO$_3$ around 200, 450 and 700 cm$^{-1}$. (c) Temperature dependence of the resistivity (black solid line) and the magnetic diffraction peak intensity (blue circles) [4] for the SmNiO$_3$ thin film. SmNiO$_3$ is a paramagnetic (PM) metal above 370 K, a charge-ordered (CO) paramagnetic insulator at 200 K < T < 370 K, and a charge-ordered antiferromagnetic (AFM) insulator below 200 K.

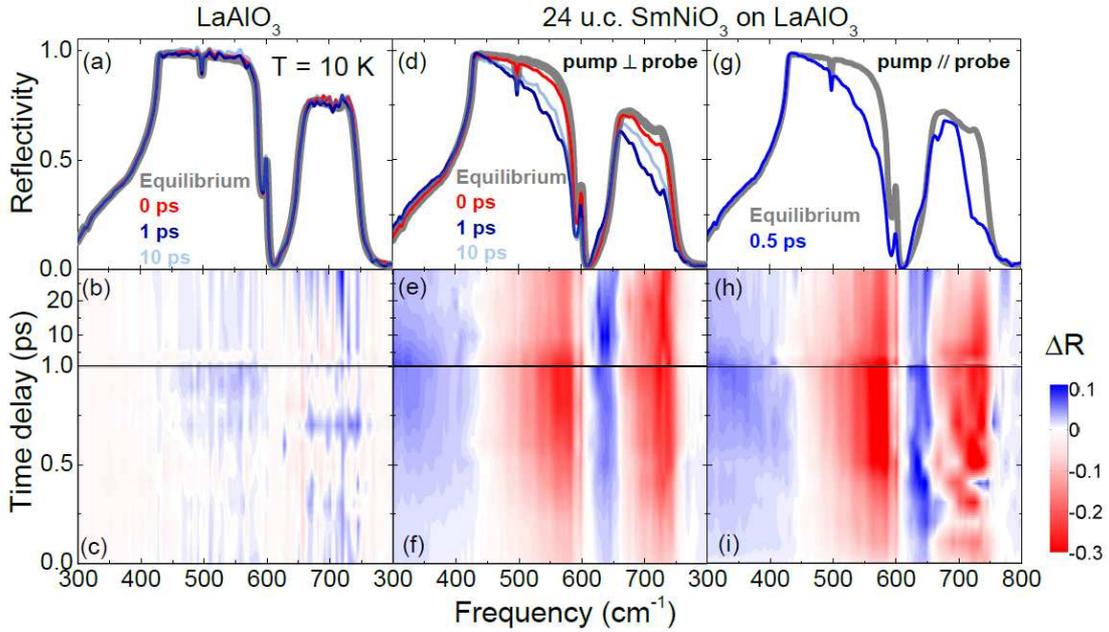

FIG. 2. (Color online) Transient optical reflectivity of the bare substrate (a-c), and the nickelate heterostructure measured with perpendicular (d-f), and parallel (g-i) pump-probe geometries at $T$ = 10 K. Upper panel: transient reflectivity at selected time delays. Middle and lower panels: transient reflectivity change from 0 to 30 ps. The 0 ps time delay is defined as the peak of the pump pulse.

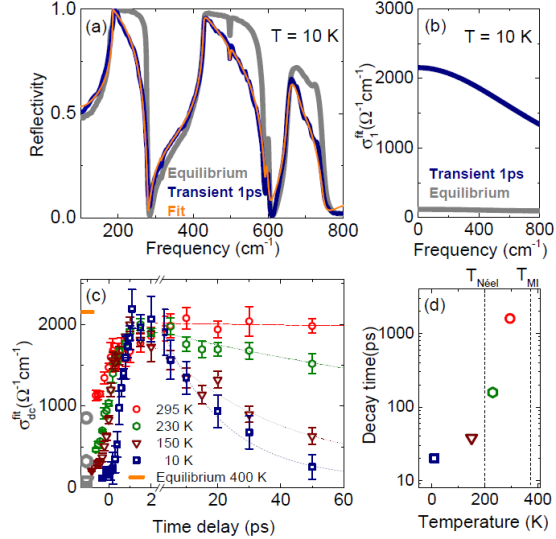

FIG. 3. (Color online) (a) Equilibrium reflectivity (grey), transient reflectivity at 1 ps (dark blue) and the fit (orange) of $SmNiO_3/LaAlO_3$. (b) The real part of the optical conductivity extracted from the fit. (c) Within the experimental error, the same maximum transient conductivity is reached for all measured temperatures. (d) The lifetime of the transient state tends to diverge approaching the metal-insulator transition temperature $T_{MI}$.

[1] J. Liu *et al.*, Nat. Commun. 4, 2714 (2013); M. Hepting *et al.*, Phys. Rev. Lett. 113, 227206 (2014); A. J. Hauser *et al.*, Appl. Phys. Lett. 106, 092104 (2015).
[2] S. Catalano *et al.*, APL Materials 2, 116110 (2014).
[3] A. D. Caviglia *et al.*, Phys. Rev. Lett. 108, 136801 (2012).
[4] M. Först *et al.*, Nature Mater. 14, 883 (2015).
[5] M. K. Stewart *et al.*, Phys. Rev. Lett. 107, 176401 (2011).
[6] Z. M. Zhang *et al.*, J. Opt. Soc. Am. B 11, 2252 (1994).
[7] G. Catalan, Phase Transit. 81, 729 (2008); M. L. Medarde, J. Phys.: Condens. Matter 9, 1679 (1997).
[8] M. V. Abrashev *et al.*, Phys. Rev. B 59, 4146 (1999).
[9] J. Ruppen *et al.*, Phys. Rev. B 92, 155145 (2015).
[10] R. Jaramillo *et al.*, Nature Phys. 10, 304 (2014).
[11] Since commercial $LaAlO_3$ samples are twined, both the bare substrate and the nickelate heterostructure show isotropic optical response.
[12] I.-C. Ho, X. Guo, and X.-C. Zhang, Opt. Express 18, 2872 (2010).
[13] R. Huber *et al.*, App. Phys.Lett. 76, 3191 (2000).
[14] W. Zhong, R. D. King-Smith, and D. Vanderbilt, Phys. Rev. Lett. 72, 3618 (1994).
[15] M. Dressel and G. Grüner, Electrodynamics of Solids, Cambridge University Press, Cambridge (2002).